\documentclass[aps,prl,twocolumn,showpacs,superscriptaddress]{revtex4-1}
\usepackage{amsfonts}

\usepackage{graphicx}
\usepackage{subfigure}
\usepackage{braket}
\usepackage{amsmath}
\usepackage{endnotes}
\usepackage{bm}

\begin{document}

\title{Strong Field Ionization as an Inhomogeneous Schr\"odinger Equation}

\author{Zachary~B.~Walters}
\affiliation{Max Planck Institute for Physics of Complex Systems, N\"othnitzer
Strasse 38, D-01187 Dresden, Germany}
\author{Jan-Michael Rost}
\affiliation{Max Planck Institute for Physics of Complex Systems, N\"othnitzer
Strasse 38, D-01187 Dresden, Germany}
\date{\today}

\begin{abstract}
Strong field ionization is difficult to treat theoretically due to the
simultaneous need to treat bound state dynamics accurately and continuum
dynamics efficiently.  We address this problem by decomposing the time
dependent Schr\"odinger equation (TDSE) into an inhomogeneous form, in which a
precomputed bound state acts as a source term for a time dependent tunneling
component.  The resulting theory is equivalent to the full TDSE when exact
propagation is used, and reduces or eliminates a major source of wavefunction
error when propagation is approximated.  The gauge invariance of the resulting
theory is used to clarify an apparent gauge dependence which has long been
observed in the context of strong field S-matrix theory.
\end{abstract}

\maketitle

The ionization of an atom or molecule by an intense laser field is one of the
basic phenomena of strong field physics, and also one of the most difficult to
describe theoretically.  The difficulty arises from the very different
theoretical techniques needed to accurately describe the bound states of the
molecule on one hand, and to efficiently describe the dynamics of the ionized
electron on the other.  In the zero field limit, molecular states can be well
described using the methods of quantum chemistry.  However, such methods
typically scale poorly with system size, making them ill suited to problems in
which one electron may travel dozens or hundreds of atomic units from the
ionic center.  Meanwhile, the dynamics of the ionized electron can be well
described using single active electron calculations, at the cost of giving a
poor description of the bound states.  In the tunneling regime, the continuum
component is typically exponentially suppressed with respect to the bound
component, so that even slight errors in the treatment of the initial state
have the potential to swamp the desired tunneling wavepacket.  

To date, most analytical methods of strong field ionization are built
around a single active electron picture, including S-matrix
theory\cite{becker2005intense} and its
variants\cite{ivanov2005anatomy,smirnova2007anatomy}, including PPT
theory \cite{perelomov1966ionization} and strong field eikonal-Volkov
\cite{smirnova2008analytical, torlina2012time}.  Methods built around
quantum chemistry approaches include multiconfiguration time dependent
Hartree and Hartree Fock \cite{nest2005multiconfiguration,
  caillat2005correlated, zanghellini2004testing} and time dependent
R-matrix approaches
\cite{moore2011rmt,lysaght2009time,moore2011time,tao2012photo}.
Coupled channel approaches \cite{walters2010attosecond,
  torlina2012timeb, spanner2009one} allow interactions with the
departing electron to drive transitions between different states of
the ion, so that the electron ``hole'' may occupy any of several
orbitals.

This paper addresses the problem of strong field ionization by
separating the evolution of a time dependent ``tunneling'' component
of the wavefunction from that of an initial bound state.  Because the
time propagator does not act upon the initial state, it may be
approximated without fear of erroneously projecting the initial state
into the continuum.  Likewise, the initial state may be found using
any desired level of theoretical approximation without increasing the
computational cost of propagation.  The resulting general method
reproduces either the full time dependent Schr\"odinger equation
(TDSE) or strong field S-matrix theory in the appropriate limits, and
is manifestly gauge invariant.  By examining its transformation under
a change of gauges, we clarify a longstanding apparent gauge
dependence in S-matrix theory.

\section{The Inhomogeneous Schr\"odinger Equation}
When subjected to an intense laser field, an electronic wavefunction begins to
deviate from its field free state.  If the initial wavefunction is an
eigenstate of the field free Hamiltonian
$\psi_{b}(x,t)=\psi_{b}(x)e^{-iE_{b}t}$, where $H_{0}\psi_{b}=E_{b}\psi_{b}$,
the full wavefunction may be written as the sum of this term plus a time
dependent ``tunneling'' component (despite the name, the logic works equally
well in the multiphoton regime)
\begin{equation}
\psi(t)=\psi_{b}(t)+ \psi_{t}(t).
\end{equation}
Mathematically, the tunneling component serves to cancel the nonzero action
residual 
\begin{equation}
\varphi(x,t)=(i\frac{\partial}{\partial t}-H(t))\phi_{b}(x,t)
\end{equation}
which results when the field-free eigenstate is acted upon by the field-on
Hamiltonian.  Writing the full TDSE as
\begin{equation}
i\frac{\partial}{\partial t} (\psi_{0}e^{-iE_{0}t}+ \psi_{t}(t))-
(H_{0}+\delta H(t))(\psi_{0}e^{-iE_{0}t}+ \psi_{t}(t))=0
\label{eq:TDSE}
\end{equation}
and subtracting the evolution of the bound state due to the field free
Hamiltonian
\begin{equation}
i\partial/\partial
t\psi_{0}e^{-iE_{0}t}-H_{0}\psi_{0}e^{-iE_{0}t}=0
\end{equation}
yields an inhomogeneous equation in which the laser field $H(t)-H_{0}$ acts
upon the bound state to serve as a source term for the tunneling component
\begin{equation}
i \frac{\partial}{\partial t} 
\psi_{t}(t)-H(t)\psi_{t}(t)=(H(t)-H_{0})\psi_{0}e^{-iE_{0}t}.
\label{eq:TDSE_inhomogeneous}
\end{equation}

Although Eqs. \ref{eq:TDSE} and \ref{eq:TDSE_inhomogeneous} are formally
equivalent when the time propagation is exact, they will behave differently
when it is approximated.  In Eq. \ref{eq:TDSE}, the propagator acts on both
the bound and the tunneling components, while in
Eq. \ref{eq:TDSE_inhomogeneous} it acts on the tunneling component alone.

If the Hamiltonian $H(t)$ is approximated by $\tilde{H}(t)$, where
$\tilde{H}(t)=H(t)-\zeta(t)$, the Hamiltonian error $\zeta(t)$ will cause the
propagated wavefunction to deviate from the true wavefunction.  Writing the
true tunneling component $\psi_{t}(x,t)=\phi_{t}(x,t)+\epsilon(x,t)$ as the
sum of the propagated component plus an error term, then propagating
$\phi_{t}$ using the approximate Hamiltonian, so that
\begin{equation}
(i\frac{\partial}{\partial t}-\tilde{H})(\psi_{b}(x,t)+\phi_{t}(x,t))=0,
\end{equation}
causes the error term to obey an inhomogeneous Schr\"odinger equation
\begin{equation}
i\frac{\partial}{\partial t}\epsilon(t)-H(t)\epsilon(t)=
\zeta(t)(\psi_{b}(x,t)+\phi_{t}(t)),
\label{eq:error_hom}
\end{equation}
where the Hamiltonian error $\zeta(t)$ acts upon both the bound and the
tunneling components of the propagated wavefunction to serve as a source term
for the error.  If the tunneling component is small relative to the bound
state component, the dominant source of error is the Hamiltonian error acting
upon the bound state.  Thus, a small error requires that $\psi_{b}$, an
eigenstate of the true Hamiltonian $H(t)$, be accurately described by the
approximate Hamiltonian $\tilde{H}(t)$ used in the propagation.

In contrast, Eq. \ref{eq:TDSE_inhomogeneous} allows for two approximations to
the Hamiltonian -- the approximation $\tilde{H}_{B}=H_{0}-\zeta_{B}$ used to
find $\phi_{B}$, and the approximation $\tilde{H}_{t}(t)=H(t)-\zeta_{t}(t)$
used to propagate $\phi_{t}$.  In terms of these Hamiltonians, the
wavefunction error now obeys
 \begin{equation}
i\frac{\partial}{\partial t} \epsilon_{t}(t)-H(t)\epsilon_{t}(t)=
\zeta_{B}\psi_{B}(x,t)+\zeta_{T}(t)\psi_{t},
\label{eq:error_inhom}
\end{equation}
so that the error in the bound state Hamiltonian operates on the bound state,
and the error in the propagation Hamiltonian acts on the tunneling component.
The advantage of the inhomogeneous TDSE, then, is the freedom to choose a
different (presumably better) approximation to the Hamiltonian when finding
the initial bound state than will subsequently be used in the propagation.
Doing so will reduce or eliminate the major source of error in
Eq. \ref{eq:error_hom} at the cost of finding a single, static eigenfunction
of $\tilde{H}_{B}$, while leaving the cost of propagation essentially
unchanged.


\subsection{Gauge Transformation}
\label{subsection:gaugetransformation}
The use of an inhomogeneous Schr\"odinger equation has a long history in
analytical approaches to strong field ionization, in particular strong field
S-matrix theory \cite{ivanov2005anatomy,smirnova2007anatomy}.  Here, the
molecular potential is often ignored completely after the moment of
ionization, so that the tunneling component can be expanded in a basis of
Volkov states $\phi_{t}=\int dp C_{p}\ket{p-eA(t)}$, where $p$ is the
conjugate momentum of the electron in the laser field, $e=-1$ is the charge of
the electron, and
\begin{equation}
C_{p}=-i\int_{0}^{t}dt' e^{-i
  \frac{1}{2}(p-eA(t''))^{2}dt''}\bra{p-eA(t)}V_{L}(t')\ket{\phi_{b}}e^{-i E_{b}t'}
\end{equation}
or, in differential form,
\begin{equation}
  (i\frac{\partial}{\partial t}-\frac{1}{2}(p-eA(t))^{2})C_{p}=
  e^{-ieA(t')x}\bra{p}V_{L}(t')\ket{\phi_{b}}
  e^{-i E_{b}t'}.
\label{eq:sfa_differential}
\end{equation}

An apparent gauge dependence now arises from the question of what gauge
$V_{L}(t)=H(t)-H_{0}$ is to be calculated in.  As the Volkov states are
stationary in the velocity gauge, it may seem natural to calculate $V_{L}$ in
this gauge as well.  However, comparisons with the numerical TDSE have shown
\cite{becker2009gauge} better agreement when $V_{L}$ is calculated in the
length gauge.  Although perfect gauge invariance cannot be expected when
ignoring the molecular potential, such discrepancies have been seen as well in
the ionization of negatively charged ions, where the strong field
approximation is better justified.  As gauge invariance is an essential
feature of a physical theory, a brief digression is warranted.



As written, Eqs. \ref{eq:TDSE} and \ref{eq:TDSE_inhomogeneous} are manifestly
gauge invariant.  Rewriting the right hand side of
Eq. \ref{eq:TDSE_inhomogeneous} as 
\begin{equation}
(H-H_{0})\psi_{0}e^{-iE_{0}t}\rightarrow(i\frac{\partial}{\partial
  t}-H(t))\psi_{0}e^{-iE_{0}t},
\end{equation}
Eq. \ref{eq:TDSE_inhomogeneous} becomes
\begin{equation}
i \frac{\partial}{\partial t} 
\psi_{t}(t)-H(t)\psi_{t}(t)=\varphi(x,t)=(i\frac{\partial}{\partial
  t}-H(t))\psi_{0}e^{-iE_{0}t},
\label{eq:TDSE_inhomogeneous_twosided}
\end{equation}
so that Eq. \ref{eq:TDSE} and both sides of
Eq. \ref{eq:TDSE_inhomogeneous_twosided} have the form
\begin{equation}
\varphi(x,t)=(i\frac{\partial}{\partial t}-H(t))\psi(x,t),
\end{equation}
where $\varphi(x,t)$ can be thought of as an action residual which
arises when $\psi(x,t)$ does not satisfy the Schr\"odinger equation
perfectly.  If we now perform the gauge transformation
$\psi(x,t)\rightarrow \psi(x,t)e^{i \Lambda(x,t)}$,
$\vec{A}(x,t)\rightarrow \vec{A}(x,t)-\vec{\nabla} \Lambda(x,t)$,
$\phi(x,t)\rightarrow \phi(x,t) + \frac{\partial}{\partial
  t}\Lambda(x,t)$, then transforming the action residual according to
$\varphi(x,t) \rightarrow \varphi(x,t) e^{i \Lambda(x,t)}$ preserves
equality.  In Eq. \ref{eq:TDSE}, $\varphi(x,t)=0$ is unaffected by the
change of gauge, while in Eq. \ref{eq:TDSE_inhomogeneous_twosided},
$\varphi(x,t)$ is affected in the same way by transforming both sides
of the equation, so that the theory as a whole is gauge invariant.

The apparent gauge dependence which is seen in strong field S-matrix theory
arises from calculating the source term in one gauge and the tunneling
component in another.  Thus, it is equivalent to transforming one side of
Eq. \ref{eq:TDSE_inhomogeneous} but not the other.  It is therefore not a true
gauge transformation which causes the observed results, but rather a subtle
modification of the effective source term.  If we reverse the one sided
transformation, so that both sides are calculated in the same gauge, the
effective source term will not be the field free bound state $\psi_{b}(t)$,
but rather $\psi_{b}(t)e^{-i\Lambda(x,t)}$.  For the specific case of strong
field S-matrix theory, the action residual from Eq. \ref{eq:sfa_differential}
becomes
\begin{equation}
\begin{split}
\varphi(x,t)=&e^{-ie\vec{A}(t)\cdot\vec{x}}\vec{E}(t)\cdot\vec{x}\psi_{b}(x)e^{-i E_{b}t'}\\ 
\rightarrow& 
(\frac{(\vec{p}-e\vec{A}(t))^{2}}{2}-\frac{|\vec{p}|^{2}}{2})\psi_{b}(x)e^{-i E_{b}t'}e^{-i
  e\vec{A}(t)\cdot\vec{x}},
\end{split}
\end{equation}
so that the effective source term when calculated in the same gauge as the
tunneling component becomes $\psi_{b}(t)e^{-i E_{b}t'}e^{-ie\vec{A}(t)\cdot\vec{x}}$.  The
improved accuracy of the ``length gauge'' calculation can now be explained by
noting that the effective source wavefunction
\begin{equation}
\psi_{b}(x,t)=\psi_{0}e^{-E_{b}t}e^{-ie\vec{A}(t)\cdot\vec{x}}
\end{equation}
obeys
\begin{equation}
i\frac{\partial}{\partial t}\psi_{b}(x,t)=
(H_{0}+\vec{E}(t)\cdot\vec{x})\psi_{b}(x,t),
\end{equation}
so that it is a quasistatic eigenfunction of the field-on Hamiltonian rather
than a static eigenfunction of the field-off Hamiltonian. By incorporating the
dressing of the bound state by the laser field into the source term rather
than the tunneling component, a correspondingly smaller fraction of the
dynamics must be treated using the inexact propagation Hamiltonian, giving a
simple explanation for the improved accuracy.

\section{Numerical Propagation}
\subsection{Least Action Propagator}
The inhomogeneous form of the Schr\"odinger Equation may be propagated using
methods similar to those used for the homogeneous form.  For this paper, we
propagated both forms using the least action propagator
\cite{walters2011efficient}.  For each timestep, we expanded both the bound
and the tunneling components of the wavefunction with respect to a spacial
basis $\chi_{i}(x)$ and a temporal basis $T_{n}(t)$, so that
$\phi(x,t)=\phi_{0}(x,t)+\phi_{t}(x,t)=\sum_{i,n}D_{i,n}\chi_{i}(x)T_{n}(t)$.
The expansion coefficients $D_{in}$ were found by minimizing the action
accumulated by the wavefunction over the chosen timestep.  By choosing finite
bases in space and time, this can be accomplished by solving a linear system
of equations for $D_{in}$.  In order to reduce computational cost, our spatial
basis was chosen to yield banded matrices, while our temporal basis was
constructed to allow for eigenvector decomposition.

For a time dependent Hamiltonian $\tilde{H}(t)$ and temporal interval
$[t_{0},t_{0}+\Delta t]$, the action accumulated over the timestep is
minimized when
\begin{equation}
(i O_{ij}Q_{nm}-\tilde{H}_{ijnm})D_{jm}=0 
\label{eq:leastaction}
\end{equation}
for all $i,n$ such that $D_{in}$ is a free parameter (some values of $D_{in}$
will be fixed by the choice of initial or boundary conditions).  Here
$O_{ij}=\braket{\chi_{i}|\chi_{j}}$ is the spatial overlap matrix,
$Q_{ij}=\bra{T_{n}}\frac{\partial}{\partial t}\ket{T_{m}}$ is the temporal
derivative matrix, and
$\tilde{H}_{ijnm}=\bra{\chi_{i}T_{n}}\tilde{H}(t)\ket{\chi_{j}T_{m}}$ is the
(time dependent) Hamiltonian matrix.  For a static Hamiltonian,
$\tilde{H}_{ijnm}=\tilde{H}_{ij}U_{nm}$, where $U_{nm}=\braket{T_{n}|T_{m}}$
is the temporal overlap matrix and
$\tilde{H}_{ij}=\bra{\chi_{i}}\tilde{H}\ket{\chi_{j}}$ is the Hamiltonian
matrix.

Separating the bound and tunneling components as in Eq
\ref{eq:TDSE_inhomogeneous} yields a linear system with a nonzero right hand
side.  In action language, the laser field causes the bound state to
accumulate nonzero action, which must be canceled by the evolution of the
tunneling component.  Writing the tunneling component as
$\phi_{t}(x,t)=\sum_{i,n}C_{in}\chi_{i}(x)T_{n}(t)$ and the field free bound state
component as $\phi_{0}=\sum_{i,n}B_{in}\chi_{i}(x)T_{n}(t)$, the least action
equation becomes
\begin{equation}
(i O_{ij}Q_{nm}-\tilde{H}_{ijnm})C_{jm}=\Delta \tilde{H}_{ijnm}B_{jm},
\label{eq:ls_inhom}
\end{equation}
where $\Delta \tilde{H}_{ijnm}=\bra{\chi_{i}T_{n}}\Delta
\tilde{H}_{B}(t)\ket{\chi_{j}T_{m}}$.

In the velocity gauge, the kinetic energy term of the Hamiltonian is given by
$\tilde{H}(t)=\frac{1}{2}(\vec{p}-e\vec{A}(t))^{2}$, where $A(t)$ is the time
dependent vector potential, with matrix elements
$A_{nm}=\bra{T_{n}}A(t)\ket{T_{m}}$ and
$A^{2}_{nm}=\bra{T_{n}}A^{2}(t)\ket{T_{m}}$ for its square.  The squared term
is often neglected in numerical treatments, but is required here for gauge
invariance and for use with exterior complex scaling.  The least action linear
system is now
\begin{equation}
\begin{split}
[i
O_{ij}Q_{nm}-(-\triangle_{ij}U_{nm}-\frac{i}{2}\nabla_{ij}A_{nm}&
\\+\frac{i}{2}\nabla^{\dag}_{ij}A_{nm}
+\frac{1}{2}O_{ij}A^{2}_{nm}+V_{ij}U_{nm}&)]C_{jm}=R_{in},
\end{split}
\label{eq:leastaction_source}
\end{equation}
where $R_{in}$ is the inhomogeneous source term,
$\nabla_{ij}=\bra{\chi_{i}}\nabla\ket{\chi_{j}}$ is the matrix element of the gradient and $\triangle_{ij}=\bra{\chi_{i}}\triangle\ket{\chi_{j}}$ is the matrix
element of the Laplacian.  As discussed earlier, the gauge in which
$R_{in}$ is found may be varied, with a corresponding change in the
effective bound state.  For this paper, we calculated $R_{in}$ in both
the velocity gauge
\begin{equation}
R_{in}=(-\frac{1}{2}i\nabla_{ij}A_{nm}+
\frac{1}{2}i\nabla^{\dag}_{ij}A_{nm}+ \frac{1}{2}O_{ij}A^{2}_{nm})B_{jm}
\end{equation}
and the length gauge
\begin{equation}
O_{ij}U_{nm}R_{jm}=\braket{\chi_{i}(x)T_{n}(t)|e^{-iA(t)x}x\phi_{b}(x,t)}.
\end{equation}

The linear system constructed in this way requires a large number of
coefficients to be found at every timestep -- for a spatial basis set of size
$N_{x}$ and a temporal basis set of size $N_{t}$, $N_{x}\cdot N_{t}$
coefficients must be found at every timestep.  We chose our spatial and
temporal basis sets so as to decompose this problem into the iterated solution
of $(N_{t}-1)$ sparse, banded linear systems for $N_{x}$ coefficients apiece,
so that every timestep required computational effort on the order of
$N_{x}N_{t}$.

\subsection{Spatial basis}
Our spatial basis was constructed following \cite{scrinzi2010infinite} to
allow for infinite range exterior complex scaling.  For a one dimensional
problem, the region $-\infty<x<\infty$ is divided into a number of finite
elements, plus two semi-infinite end caps.  Within each element, the
wavefunction is described using a primitive basis of low order Legendre
polynomials.  In each end cap, the primitive basis set is given by Laguerre
polynomials times declining exponentials, so that the wavefunction is required
to be 0 at $x=\pm\infty$.  In contrast to \cite{scrinzi2010infinite}, here the
same order polynomials were used in each element and both end caps.

We dealt with the problem of reflections from the borders of the propagation
region by employing infinite range exterior complex scaling.  Beyond some
scaling radius $x_{ecs}$, we analytically rotated $x\rightarrow
x_{ecs}+(x-x_{ecs})e^{-i\theta}$, so that a wavefunction of the form $e^{ikx}$
will go to zero as $x$ goes to infinity.  This was accomplished by multiplying
$O_{ij} \rightarrow O_{ij}e^{-i\theta}$,
$\triangle_{ij}\rightarrow\triangle_{ij}e^{i\theta}$, $V_{ij}\rightarrow
V_{ij}e^{-i\theta}$ for all elements beyond $x_{ecs}$.  Because exterior
complex scaling causes a derivative discontinuity at $x_{ecs}$, it is
necessary to enforce an element boundary at $x_{ecs}$.  In contrast to
\cite{scrinzi2010infinite}, we include a number of finite elements in the
scaled region in addition to the semi-infinite end caps.  The inner boundary
$x_{b}$ of the end caps was chosen such that $e^{-k|x_{b}-x_{ecs}|}$, where
$k=\sqrt{2 E_{max}}$, was less than some tolerance.

Aside from enforced element boundaries at $x=0$ and $x=\pm x_{ecs}$, the size
of an individual element was determined using a WKB approach.  Choosing an
energy cutoff $E_{max}$ and a maximum phase $\Delta \phi$ to be accumulated in
an element, the size $\Delta x$ of an element was given by $\Delta \phi=\Delta
x\sqrt{2(E_{max}+|V(x)|+|Fx|)}$, where $F$ is the maximum strength of the
laser field.  Constructing the spatial basis in this way ensures that a local
Hamiltonian will yield sparse, banded spatial matrices, with a bandwidth
determined by the order of the polynomials in the primitive basis set.  Highly
oscillatory functions can be treated by decreasing the size of the finite
elements, or by increasing the order of the polynomials used inside each
element.

Wavefunction continuity across element boundaries was enforced by constructing
linear combinations of the primitive basis sets.  Within each element, we
constructed left border functions $BF_{L}(y)=\frac{1}{2}(P_{0}(y)-P_{1}(y))$,
right border functions $BF_{R}(y)=\frac{1}{2}(P_{0}(y)+P_{1}(y))$, and
interior functions $BF_{n}(y)=(P_{n}(y)-P_{n \mod 2}(y))$ for $n\ge 2$, where
$y$ is a local coordinate equal to $-1$ at the left boundary and $1$ at the
right boundary.  Recalling that $P_{l}(-1)=(-1)^{l}$ and $P_{l}(1)=1$, it can
be seen that $BF_{L}$ is the only nonzero function at the left boundary and
$BF_{R}$ is the only nonzero function on the right boundary.  Wavefunction
continuity was enforced by requiring the temporal coefficients $C_{in}$ of the
left border function in one element to equal the coefficients of the right
border function in the element to the left, and vice versa.  Border functions
in the end caps were defined in an analogous way using Laguerre rather than
Legendre polynomials.

\subsection{Temporal basis}
While the spatial basis was constructed to yield sparse, banded spatial
matrices and to enforce wavefunction continuity, the temporal basis set was
constructed to reduce the size of the linear systems which must be solved at
every timestep.  Linear combinations of the primitive Legendre basis were
constructed so that only one basis function was nonzero at the beginning of
the timestep, while the other basis functions had a tridiagonal overlap
structure.  Using local coordinate $\tau$, equal to $-1$ at the beginning of
the timestep and $+1$ at the end, we set
$T_{0}(t)=\frac{1}{2}(P_{0}(\tau)-P_{1}(\tau))$ and
$T_{n}(t)=\frac{1}{2}(P_{n}(\tau)+P_{n-1}(\tau))$, where $P_{n}(\tau)$ is the
Legendre polynomial of order n.  Note that $T_{0}(t_{0})=1$,
$T_{0}(t_{0}+\Delta t)=0$, and  $T_{n}(t_{0})=0$, $T_{n}(t_{0}+\Delta t)=1$
for $n\ge 1$.  Thus, the wavefunction at the beginning of the step is given by
the coefficients of the zeroth order basis function, while the wavefunction at
the end of the pulse is given by the sum of the coefficients of all the
nonzero basis functions.

The size of the linear systems to be solved can be reduced by performing an
eigenvector decomposition of the residuals $R_{in}$, $n\ge 1$, which must be
eliminated by the tunneling wavefunction.  (Because $C_{i0}$ are not free
parameters, $R_{i0}$ will not in general be canceled by the minimum action
solution.)  Noting that for small timesteps $A(t)$ and $A^{2}(t)$ are
approximately constant, so that $A_{nm}\approx a U_{nm}$ and $A^{2}_{nm}
\approx b U_{nm}$, the residuals can be projected onto generalized
eigenvectors of the $Q$ and $U$ matrices.  Setting $\tilde{Q}_{nm}=Q_{nm}$ and $\tilde{U}_{nm}=U_{nm}$ for $n,m \ge 1$,
$\tilde{U}$ and $\tilde{Q}$ have right and left eigenvectors
\begin{equation}
\tilde{Q}v^{\alpha}=\lambda^{\alpha}\tilde{U}v^{\alpha}
\end{equation}
and
\begin{equation}
w^{\alpha}\tilde{Q}=\lambda^{\alpha}w^{\alpha}\tilde{U},
\end{equation}
where $\lambda^{\alpha}$ is a generalized eigenvalue and
$w^{\alpha\dag}\tilde{U}v^{\beta}=\delta_{\alpha\beta}$.  Ignoring
$(A_{nm}-aU_{nm})$ and $(A^{2}_{nm}-bU_{nm})$ for the moment, if $\delta
C_{jm}=\sum_{\alpha}C^{\alpha}_{j}v^{\alpha}_{m}$, then
\begin{equation}
(i O_{ij}\lambda^{\alpha}-(-\nabla^{2}_{ij}-ia\nabla_{ij}+
ia\nabla^{\dag}_{ij}+bO_{ij}+V_{ij}))\delta C^{\alpha}_{j}=
\sum_{n}w^{\alpha\dag}_{n}R_{in}.
\end{equation}
Because this method neglects the differences $(A_{nm}-aU_{nm})$ and
$(A^{2}_{nm}-bU_{nm})$, it will yield an approximate rather than exact
solution for $\delta C^{\alpha}_{j}$, so that applying the correction
$C_{in}\rightarrow C_{in}+\delta C_{in}$ will not cancel the residuals
completely.  However, as a single timestep will typically be very short
compared to the frequency of the driving laser, the variation of $A(t)$ and
$A^{2}(t)$ over a single timestep will be very small, and repeated application
of this procedure will yield rapid convergence to zero residual.  We repeat
this iteration until $\delta \psi_{t}(x,t+\Delta t)=\sum_{n}\delta C_{in}$ has
a norm less than some desired tolerance.  After finding the minimum norm
solution, we adjust the size of the next step so that the contribution of the
last temporal basis function to the final wavefunction $\delta
\psi_{n_{max}}=\sum_{i}C_{in_{max}}\chi_{i}(x)$ has a norm less than a chosen
accuracy goal.

Having found the variationally optimum description of the wavefunction in the
chosen basis, the timestep may now be adjusted to ensure that the wavefunction
evolution is well described by the temporal basis.  We ensure this by choosing
a timestep such that the last Legendre polynomial included in the primitive
basis has a negligible contribution to the total solution.  If
\begin{equation}
F_{i}=U^{-1}\sum_{n}C_{in}\braket{P_{n_{max}}|T_{n}}
\end{equation} 
is the vector of spatial coefficients for the last Legendre polynomial and
\begin{equation}
\delta \psi_{n_{max}}=\sum_{i,j}F^{*}_{i}O_{ij}F_{j},
\end{equation}  
is its norm, then 
\begin{equation}
\frac{\Delta t'}{\Delta t}=
(\frac{\text{accuracy goal}}{\delta \psi_{n_{max}}})^{n_{max}-1}
\end{equation}
gives the ratio of the new time step to the old.

These design choices yield a variable stepsize, variationally optimum
propagator for a wavefunction described in a finite element basis.  The order
of the polynomials used to describe the wavefunction within each element, as
well as the temporal order used to describe the wavefunction over a given
timestep, can be chosen at will.  The linear systems which must be solved are
sparse, banded, and do not grow with increasing temporal order.  For the
calculations in this paper, we used polynomials of order 5 for both our
spatial and temporal bases, and propagated with an accuracy goal of
$10^{-6}$.  As the spatial component $\delta \psi_{n_{\text{max}}}$ of the last
temporal basis function vanishes very rapidly with increasing
$n_{\text{max}}$, the maximum step size was limited, not by the accuracy of
the propagator, but rather by the accuracy of a polynomial expansion for the
temporal evolution $e^{iE_{g}t}$ of the source wavefunction.

\section{Results}
We illustrate our treatment of strong field ionization by comparing the
evolution of the tunneling component to that of the full wavefunction
resulting from half a cycle of an intense laser pulse.  For our model atom we
use a one dimensional soft core potential
$V(x)=-\frac{Q}{\sqrt{a^{2}+x^{2}}}$, where $Q=a=1$, whose ground state has
energy $E_{g}=-0.670$.  For laser electric field $E(t)=F_{0}\cos(\omega t)$,
we set $F_{0}=0.05$ and $\omega=0.0565$ (equivalent to an intensity of $8.9
\times 10^{13}$ W/cm$^{2}$, consistent with many strong field experiments).
The combination of ground state and laser pulse yielded a Keldysh parameter
$\gamma=1.31$, placing it in the intermediate range between the tunneling and
multiphoton limits.

We compare the evolution of the full wavefunction to that of the tunneling
component using three different potentials for the propagator.  First, we
verify that our approach gives results identical to the homogeneous TDSE by
comparing comparing their evolution using the exact propagator.  We then
demonstrate the effects of an imperfect propagator by intentionally distorting
the potential used in propagation.  We perform one calculation using $V(x)=0$
(the well known strong field approximation) and one with
$V(x)=-\frac{Q}{\sqrt{a^{2}+x^{2}}}$ with $a\rightarrow 2.0$, so that the
potential is asymptotically correct but differs from the correct potential at
short range.  Such a distortion is analogous to those which might arise in a
single active electron mean field calculation, where the electron's
interaction with the ion is correct at long range but differs from the true
potential at short range.

\begin{figure}[h!]
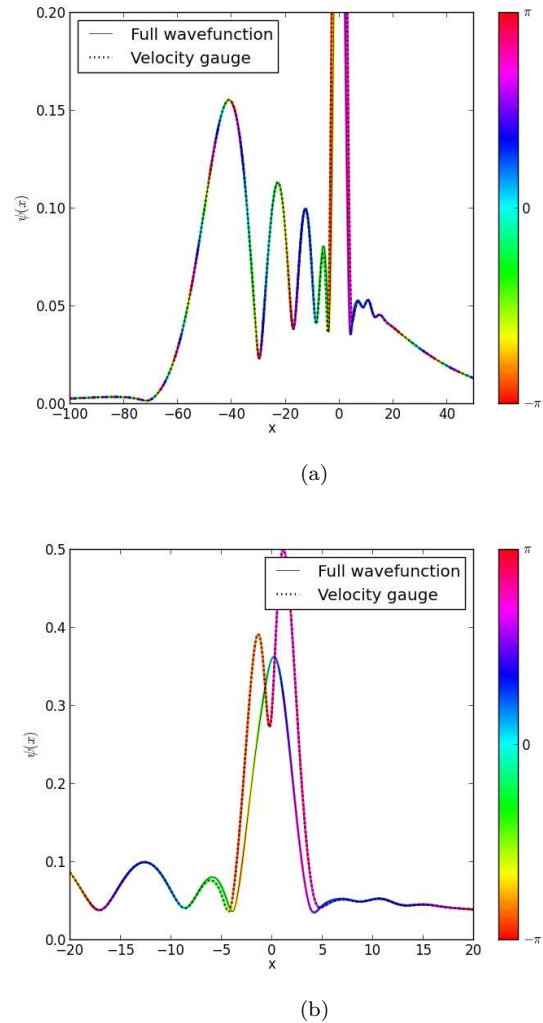

  \begin{center}
    \subfigure[]{
      \includegraphics[width=\columnwidth]{F0.05_t0_m.5pi_tf_.5pi_velocitygauge_vs_fullwf.eps2}
    }
    \subfigure[]{
      \includegraphics[width=\columnwidth]{F0.05_t0_m.5pi_tf_.5pi_velocitygauge_vs_fullwf_zoomed_in.eps2}
    }

  \end{center}
  \caption{a) Tunneling component vs full wavefunction using velocity gauge
    source term b) Same calculation near the origin.  The tunneling component
    equals the full wavefunction in the region where the source wavefunction
    is zero, and differs where it is not.  Lines show wavefunction magnitude,
    while color indicates phase.}
  \label{fig:velocitygauge_vs_fullwf}
\end{figure}

In Figures \ref{fig:velocitygauge_vs_fullwf} and
\ref{fig:lengthgauge_vs_fullwf}, we show that the inhomogeneous Schr\"odinger
equation is identical to the homogeneous form for both the velocity and length
gauge source terms.  As discussed earlier, use of the ``length gauge'' source
term is equivalent to using a quasistatic rather than a static wavefunction in
the velocity gauge source term.  In the present context, this means that the
initial condition for the full wavefunction, when $\psi_{t}(x,t_{i})=0$, must
be the static bound state $\psi(x,t_{i})=\psi_{g}e^{-iE_{g}t_{i}}$ for
comparison with the velocity gauge calculation, and the quasistatic state
$\psi(x,t_{i})=\psi_{g}e^{-iE_{g}t_{i}}e^{-i\vec{A}(x,t)\cdot\vec{x}}$ for
comparison with the length gauge.  As seen in these figures, both source
terms give results which are identical to the full wavefunction in the region
where $\psi_{g}(x)=0$.  The apparent gauge dependence is thus reconciled: both
gauges give results which are equivalent to the full TDSE, and differ from
each other only due to the use of different source terms.

\begin{figure}[h!]
  \begin{center}
    \subfigure[]{
      \includegraphics[width=\columnwidth]{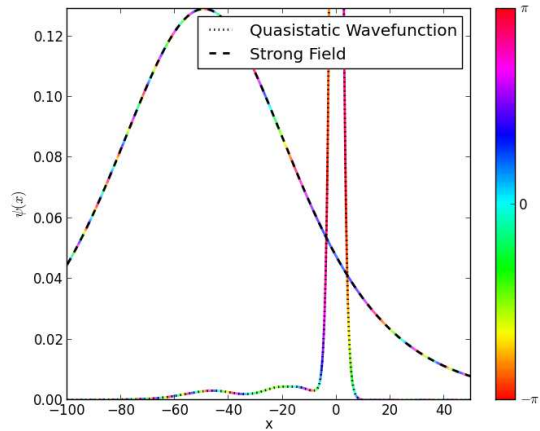}
    }
    \subfigure[]{
      \includegraphics[width=\columnwidth]{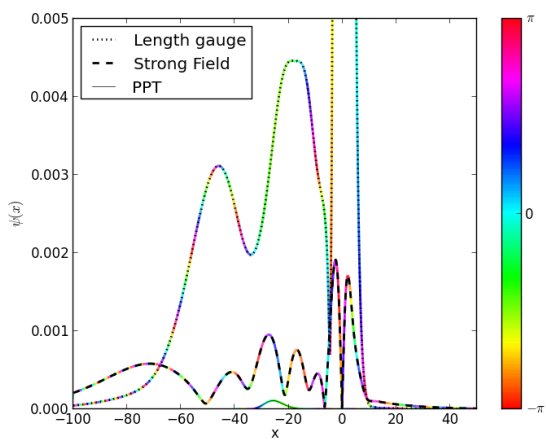}
    }
  \end{center}
  \caption{a) Full wavefunction propagated using exact propagator vs strong
    field propagator for initially quasistatic state b) Length gauge tunneling
    component propagated using exact propagator vs strong field propagator and
    PPT wavepacket.  Whereas propagation of the full wavefunction fails
    completely for the strong field propagator, the tunneling component has
    comparable magnitude to that found using the exact propagator.  The PPT
    wavepacket shows a markedly different magnitude and structure from the
    full strong field calculation.  Lines show wavefunction magnitude, while
    color indicates phase.}
  \label{fig:exact_vs_strongfield_lengthgauge}
\end{figure}

The strong field approximation is shown in Figures
\ref{fig:exact_vs_strongfield_velocitygauge} and
\ref{fig:exact_vs_strongfield_lengthgauge}.  Here setting $V(x)=0$ means that
the propagator is incapable of treating a bound state correctly, so that the
entire propagated wavefunction is projected into the continuum.  As might be
expected, propagating the full wavefunction gives results which bear no
resemblance to the true behavior.  In contrast, the inhomogeneous approach
allows the propagator only to act on that portion of the wavefunction which
differs from the initial bound state.  Here the tunneling components have
magnitudes similar to the true tunneling components, with peaks somewhat
displaced due to the different potentials seen by the continuum wavefunctions.
As the strong field approximation is commonly used in analytical treatments of
strong field ionization, we show in Figure
\ref{fig:exact_vs_strongfield_lengthgauge}b the PPT tunneling component taken
from \cite{ivanov2005anatomy}.  As in \cite{ivanov2005anatomy}, only the
exponential suppression due to tunneling is calculated, so that the ionization
amplitude has only exponential accuracy.  The PPT wavepacket is much more
localized than either the true tunneling component or the propagated strong
field result, and has a considerably smaller amplitude.  As interactions with
the departing electron may cause dynamics within the parent ion, this may be
an indication that more sophisticated treatments of strong field ionization
will be necessary to describe processes such as multichannel ionization.

\begin{figure}[h!]
  \begin{center}
    \subfigure[]{
      \includegraphics[width=\columnwidth]{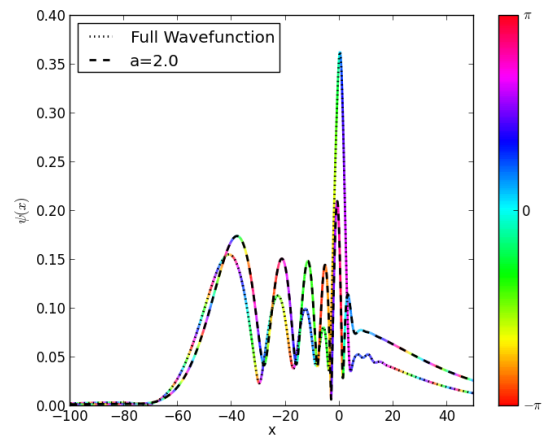}
    }
    \subfigure[]{
      \includegraphics[width=\columnwidth]{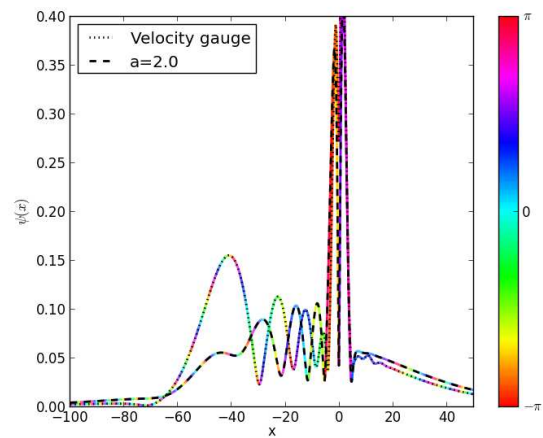}
    }
  \end{center}
  \caption{a) Full wavefunction propagated using exact propagator
    vs. propagator with modified short range potential b) Velocity gauge
    tunneling component propagated using exact propagator vs propagator with
    modified short range potential.  Lines show wavefunction magnitude, while
    color indicates phase.}
  \label{fig:exact_vs_rho_2.0_velocitygauge}
\end{figure}
\begin{figure}[h!]
  \begin{center}
    \subfigure[]{
      \includegraphics[width=\columnwidth]{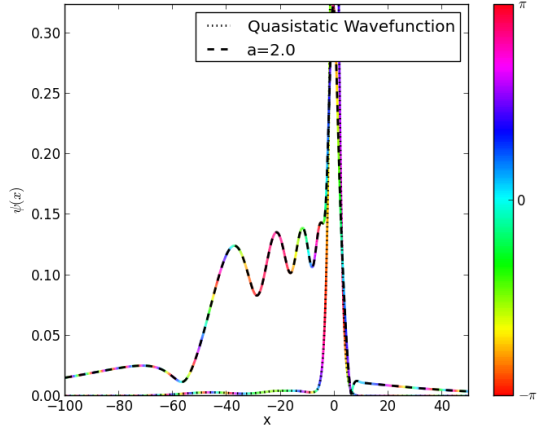}
    }
    \subfigure[]{
      \includegraphics[width=\columnwidth]{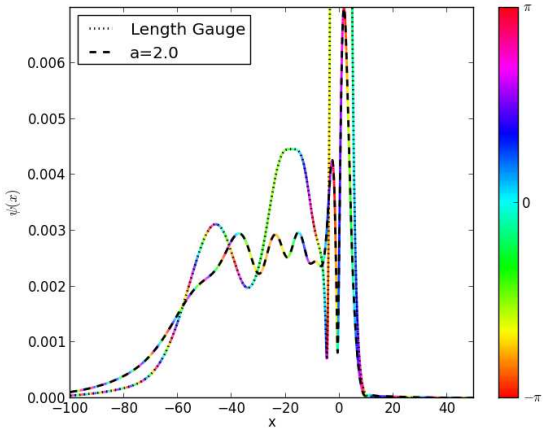}
    }
  \end{center}
  \caption{a) Full wavefunction propagated using exact propagator vs
    propagator with modified short range potential for initially quasistatic
    state b) Length gauge tunneling component propagated using exact
    propagator vs propagator with modified short range potential.  Lines show
    wavefunction magnitude, while color indicates phase. }
  \label{fig:exact_vs_rho_2.0_lengthguage}
\end{figure}

For our third calculation, we observe the results of a distorted short range
potential in Figures \ref{fig:exact_vs_rho_2.0_velocitygauge} and
\ref{fig:exact_vs_rho_2.0_lengthguage}.  In contrast to the strong field
approximation, this propagator treats the wavefunction correctly when it is
far from the molecule, but incorrectly when the electron is nearby.  Thus,
propagating the full wavefunction will cause bound states to be projected into
the continuum, but the continuum wavepacket will evolve correctly for most of
its journey.

In the velocity gauge calculation, the full wavefunction shown in Figure
\ref{fig:exact_vs_rho_2.0_velocitygauge}a has a very similar peak structure to
the exact solution, but overestimates the amplitude of the wavefunction far
from the origin.  The tunneling component in Figure
\ref{fig:exact_vs_rho_2.0_velocitygauge}b has amplitude comparable to the
exact solution, but the maxima of the tunneling component are displaced
somewhat from the exact solution.  

For the length gauge calculation, the incorrect potential causes the initially
quasistatic wavefunction to diverge strongly from the exact solution in Figure 
\ref{fig:exact_vs_rho_2.0_lengthguage}a, as a portion of the initial state is
projected into the continuum.  In contrast, Figure
\ref{fig:exact_vs_rho_2.0_lengthguage}b shows a strong similarity between the
exact and approximated tunneling components, with maxima which are once again
displaced from the exact solution.


%

\begin{figure}[h!]
  \begin{center}
    \subfigure[]{
      \includegraphics[width=\columnwidth]{F0.05_t0_m.5pi_tf_.5pi_lengthgauge_vs_quasistaticwf.eps2}
    }
    \subfigure[]{
      \includegraphics[width=\columnwidth]{F0.05_t0_m.5pi_tf_.5pi_lengthgauge_vs_quasistaticwf_zoomed_in.eps2}
    }
  \end{center}
  \caption{a) Tunneling component vs full wavefunction using length gauge
    source term b) Same calculation near the origin.  The tunneling component
    equals the full wavefunction in the region where the source wavefunction
    is zero, and differs where it is not.  The initial condition for the full
    wavefunction is the quasistatic eigenfunction rather than the field free
    eigenstate.  Lines show wavefunction magnitude, while color indicates
    phase.}
  \label{fig:lengthgauge_vs_fullwf}
\end{figure}

\begin{figure}[h!]
  \begin{center}
    \subfigure[]{
      \includegraphics[width=\columnwidth]{F0.05_t0_m.5pi_tf_.5pi_fullwf_vs_strongfield.eps2}
    }
    \subfigure[]{
      \includegraphics[width=\columnwidth]{F0.05_t0_m.5pi_tf_.5pi_velocitygauge_vs_strongfield.eps2}
    }
  \end{center}
  \caption{a) Full wavefunction propagated using exact propagator vs strong
    field propagator b) Velocity gauge tunneling component propagated using
    exact propagator vs strong field propagator.  Whereas propagation of the
    full wavefunction fails completely for the strong field propagator, the
    tunneling component has comparable magnitude to that found using the exact
    propagator.  Lines show wavefunction magnitude, while color indicates
    phase.}
  \label{fig:exact_vs_strongfield_velocitygauge}
\end{figure}

\section{Conclusions}
This paper has addressed the problem of strong field ionization from the
perspective of the time dependent Schr\"odinger equation.  Defining the
tunneling component as that part of the wavefunction which differs from a
precalculated bound state, we derived an inhomogeneous Schr\"odinger equation
which governs its evolution.  Our approach is manifestly gauge invariant, and
we have exploited this invariance to shed light on an apparent gauge
dependence which has long been puzzling in the context of the strong field
approximation.

While our approach is formally equivalent to the full time dependent
Schr\"odinger equation in the limit that an exact propagator is used, it
eliminates a major error term when the propagation is approximate.  It thus
addressed a common problem in the theory of molecular strong field dynamics,
where the computational methods necessary to describe the bound state may be
prohibitively expensive to be used in a time dependent problem.  Use of the
inhomogeneous Schr\"odinger equation may thus represent a middle ground,
allowing both the bound and the tunneling components to be treated using
theoretical approaches which are most appropriate to the task.

\end{document}